\documentclass[superscriptaddress, 12pt]{revtex4-2}
\pdfoutput=1

\usepackage[utf8]{inputenc}
\usepackage{comment}
\usepackage[english]{babel}
\usepackage{amsfonts, amsmath, amssymb, amsthm}
\usepackage[colorlinks=true,breaklinks=true,linkcolor=black,citecolor=green,urlcolor=magenta]{hyperref}
\usepackage{graphicx}

\theoremstyle{remark}

\usepackage{xcolor}
\usepackage[capitalise]{cleveref}
\usepackage{algorithmicx}
\usepackage{algorithm}
\usepackage[noend]{algpseudocode}
\usepackage[normalem]{ulem}

\algnewcommand\algorithmicinput{\textbf{Input:}}
\algnewcommand\Input{\item[\algorithmicinput]}

\algnewcommand\algorithmicoutput{\textbf{Output:}}
\algnewcommand\Output{\item[\algorithmicoutput]}

\usepackage{bm}
\usepackage{braket}

\newcommand{\nocontentsline}[3]{}
\newcommand{\tocless}[2]{\bgroup\let\addcontentsline=\nocontentsline#1{#2}\egroup}

\begin{document}

\title{Accurate quantum-centric simulations of supramolecular interactions}

\author{Danil Kaliakin}
\affiliation{Center for Computational Life Sciences, Lerner Research Institute, The Cleveland Clinic, Cleveland, Ohio 44106, United States}

\author{Akhil Shajan}
\affiliation{Center for Computational Life Sciences, Lerner Research Institute, The Cleveland Clinic, Cleveland, Ohio 44106, United States}
\affiliation{Department of Chemistry, Michigan State University, East Lansing, Michigan 48824, United States}

\author{Javier Robledo Moreno}
\affiliation{IBM Quantum, IBM T.J. Watson Research Center, Yorktown Heights, NY 10598, United States}

\author{Zhen Li}
\affiliation{Center for Computational Life Sciences, Lerner Research Institute, The Cleveland Clinic, Cleveland, Ohio 44106, United States}

\author{Abhishek Mitra}
\affiliation{Center for Computational Life Sciences, Lerner Research Institute, The Cleveland Clinic, Cleveland, Ohio 44106, United States}

\author{Mario Motta}
\affiliation{IBM Quantum, IBM T.J. Watson Research Center, Yorktown Heights, NY 10598, United States}

\author{Caleb Johnson}
\affiliation{IBM Quantum, IBM T.J. Watson Research Center, Yorktown Heights, NY 10598, United States}

\author{Abdullah Ash Saki}
\affiliation{IBM Quantum, IBM T.J. Watson Research Center, Yorktown Heights, NY 10598, United States}

\author{Susanta Das}
\affiliation{Center for Computational Life Sciences, Lerner Research Institute, The Cleveland Clinic, Cleveland, Ohio 44106, United States}

\author{Iskandar Sitdikov}
\affiliation{IBM Quantum, IBM T.J. Watson Research Center, Yorktown Heights, NY 10598, United States}

\author{Antonio Mezzacapo}
\affiliation{IBM Quantum, IBM T.J. Watson Research Center, Yorktown Heights, NY 10598, United States}

\author{Kenneth M. Merz Jr.}
\email{kmerz1@gmail.com}
\affiliation{Center for Computational Life Sciences, Lerner Research Institute, The Cleveland Clinic, Cleveland, Ohio 44106, United States}
\affiliation{Department of Chemistry, Michigan State University, East Lansing, Michigan 48824, United States}


\maketitle

\textbf{
 We present the first quantum-centric simulations of noncovalent interactions using a supramolecular approach. We simulate the potential energy surfaces (PES) of the water and methane dimers, featuring hydrophilic and hydrophobic interactions, respectively, with a sample-based quantum diagonalization (SQD) approach. Our simulations on quantum processors, using 27- and 36-qubit circuits, are in remarkable agreement with classical methods, deviating from complete active space configuration interaction (CASCI) and coupled-cluster singles, doubles, and perturbative triples (CCSD(T)) within 1 kcal/mol in the equilibrium regions of the PES. Finally, we test the capacity limits of the quantum methods for capturing hydrophobic interactions with an experiment on 54 qubits. These results mark significant progress in the application of quantum computing to chemical problems, paving the way for more accurate modeling of noncovalent interactions in complex systems critical to the biological, chemical and pharmaceutical sciences.
}
\section{Introduction}

The accurate treatment of noncovalent interactions~\cite{muller2000noncovalent,puzzarini2020challenge} is extremely important in the biological, chemical, and pharmaceutical sciences~\cite{pichler2022important,aljoundi2020covalent}. Specifically, non-covalent interactions between hydrophobic species and hydrogen-bonded pairs play pivotal roles in a myriad of biological processes, ranging from protein folding~\cite{5,6,7,8,9}membrane assembly~\cite{10}, cell signaling~\cite{11} and drug discovery ~\cite{Jhoti_Leach_2007,12,13,14,15}. The correct modeling of these interactions along with solvation plays a key role in understanding many chemical and biological processes ~\cite{hobza}.

Traditionally, quantum mechanical~\cite{24,25} methods have been used to study these systems to a high level of accuracy - so-called chemical accuracy (±1 kcal/mol from experiment). However, these calculations are quite expensive and approaches to accelerate these calculations continue to be explored using classical hardware.~\cite{Gao2024, Vogiatzis2017, Menczer2024, Ren2024, Jin2024, Schwenke2023, Delcey2023, Corzo2023, Datta2023, Zhai2021} Using the results from these calculations, force fields have been fine-tuned for wide use in molecular simulation studies of chemical and biological processes~\cite{RN49, NERENBERG2018129, RN51, RN52, RN53}. More recently, machine learning~\cite{RN50} methods built using accurate quantum chemical calculations on thousands of systems have appeared to study largely covalent interactions, but can be extended to non-covalent interactions at a good level of accuracy, but at a reduced computational cost. However, these latter methods build models that can can struggle to study diverse systems outside of the training set and can be subject to overfitting~\cite{27,28}. 

Quantum computing based studies of these interactions, to a high level of accuracy and speed, would revolutionize our ability to understand complex processes like drug binding, but would also allow for the development of large synthetic datasets that could be used to build even better force fields and quantum machine learning models. However, to date, quantum hardware has struggled to address these problems. In this work we demonstrate that quantum-centric supercomputing (QCSC)~\cite{Alexeev2024} combined with the sample-based quantum diagonalization (SQD) approach~\cite{robledo2024chemistry} allows for the study of intermolecular interactions. 

QCSC is a new computational paradigm, in which a quantum computer operates in concert with classical high-performance computing (HPC) resources. Classical processing carried out before, during, and after quantum computations allows for the introduction of quantum subroutines in the workflow of classical HPC algorithms, to extract and amplify signal from noisy quantum devices, and to leverage quantum processors to execute a limited number of large quantum circuits.

The QCSC architecture enables scaling of computational capabilities, as exemplified by methods that use classical diagonalization in subspaces determined by quantum samples such as SQD~\cite{robledo2024chemistry} and QSCI~\cite{kanno2023quantum}. The SQD method is developed based on QSCI. The SQD method use a quantum device to sample electronic configurations from a quantum circuit approximating the ground state of a molecular Hamiltonian, and use classical distributed HPC resources to post-process quantum measurements against known symmetries to obtain recovered configurations~\cite{robledo2024chemistry}, as well as to solve the Schr\"{o}dinger equation in the subspace spanned by the recovered configurations. The SQD method recently allowed us to address instances of the electronic structure problem with up to 36 spatial orbitals using up to 77 qubits~\cite{robledo2024chemistry}. The QCSC workflows
produced significant improvements over simulations using quantum computers in isolation -- which have in the last decade, used up to a handful of qubits with limited accuracies~\cite{Kandala2017,kandala2019error,google2020hartree,gao2021computational,gao2021applications,gocho2023excited,shirai2023computational,nam2020ground,khan2023chemically,obrien2023purification,zhao2023orbital,grimsley2019adaptive,christopoulou2024quantum,eddins2022doubling,motta2023quantum,castellanos2024quantum,liepuoniute2024simulation,weaving2024contextual,colless2018computation,dimitrov2023pushing,guo2024experimental,hempel2018quantum,huang2022variational,jones2024ground,kawashima2021optimizing,kirsopp2022quantum,leyton2023quantum,liang2024napa,liu2023performing,lolur2023reference,mccaskey2019quantum,rice2021quantum,santagati2018witnessing,shen2017quantum,smart2019quantum,yamamoto2022quantum,peruzzo2014variational,omalley2016scalable}. The QCSC paradigm coupled with SQD enables the study of problems heretofore out of reach of quantum computers including static correlation in iron-sulfur complexes~\cite{robledo2024chemistry} and well as dynamical correlation as exemplified in the intermolecular interactions studied herein.

Past studies have reported the simulation of noncovalent interactions~\cite{loipersberger2023accurate,ollitrault2024estimation} using symmetry-adapted perturbation theory (SAPT). This method expresses the interaction energy through a perturbative treatment of the intermolecular potential~\cite{jeziorski1994perturbation,jeziorski1976first,patkowski2020recent}, and requires the simulation of electronic structure of individual monomers on a quantum computer. In addition Anderson 
et al. demonstrated the possibility of simulations of coarse-grained intermolecular interactions on quantum computer as well.~\cite{anderson2022coarse} However, to date, predicting binding energies between monomers using the supramolecular approach, where the electronic structure of entire dimer need to be simulated on quantum hardware, has been an elusive target for quantum simulations, due to lack of accuracy and scale of conventional quantum approaches. 

Herein, we present the first quantum-centric simulation for the modeling of noncovalent hydrophilic and hydrophobic interactions with a supramolecular approach. We simulate the potential energy surfaces (PES) of the water dimer and the methane dimer. Our water dimer simulations use 27-qubit circuits, while the methane dimer simulations use 36- and 54-qubit circuits. To assess the accuracy of our quantum solutions, we compare them against heat-bath configuration interaction (HCI)~\cite{holmes2016heat,holmes2016efficient,smith2017cheap,sharma2017semistochastic} in the case of (16e,24o) calculations, complete active space configuration interaction for the (16e,12o) and (16e,16o) calculations, as well as coupled-cluster singles, doubles and perturbative triples (CCSD(T))~\cite{bartlett2024perspective} performed for all of the studied instances. The latter is widely recognized as the gold standard for computing intermolecular interactions~\cite{carsky2010recent} to chemical accuracy. For the 27-qubit water dimer and the 36-qubit methane dimer simulations, we demonstrate that SQD energies agree with CASCI nearly exactly, while deviating from CCSD(T) within 1 kcal/mol in the equilibrium region of the PES. For the 54-qubit simulations of the methane dimer, we observe how the accuracy of the quantum solution can be systematically improved by increasing the number of sampled configurations.
\section{Methods and Computational Details}

\subsection{Classical benchmark}

In the supramolecular approach binding energies between two monomers in a dimer is most often expressed as
\begin{equation}\label{eq:eq1}
    E_{\mathrm{binding}}=E_{\mathrm{AB}}-E_{\mathrm{A}}-E_{\mathrm{B}}
    \;.
\end{equation}
In Eq.~\eqref{eq:eq1} $E_{\mathrm{AB}}$, $E_{\mathrm{A}}$, and $E_{\mathrm{B}}$ denote the ground-state energies of the dimer $AB$, monomer $A$, and monomer $B$, respectively. For calculations utilizing active spaces the highest accuracy obtainable with the supramolecular approach can be achieved if Eq.~\eqref{eq:eq1} is instead expressed in terms of the energy of bound and unbound dimers ($E_{\mathrm{AB-bound}}$ and $E_{\mathrm{AB-unbound}}$). Better accuracy is achieved within this approximation due to the fact that it allows for a consistent active space in all of the calculations. Hence, in all of our calculations we express the binding energy as
\begin{equation}\label{eq2}
    E_{\mathrm{binding}}=E_{\mathrm{AB-bound}} - E_{\mathrm{AB-unbound}} \;.
\end{equation}
Here, the $E_{\mathrm{AB-unbound}}$ term of Eq.~\eqref{eq2} is approximated as two monomers separated by a 48.000 {\AA} distance, where the chosen distance guarantees the absence of interactions between the monomers.

{
\begin{table}[ht]
\caption{Active spaces used in the present work.}
\begin{tabular}{llll}
\hline\hline
species       & active space & AOs & Figure \\
\hline
water dimer   & (16e,12o) & O[2s,2p], H[1s] & \ref{fig: Figure 1}a \\
methane dimer & (16e,16o) & C[2s,2p], H[1s] & \ref{fig: Figure 1}b \\
methane dimer & (16e,24o)  & C[2s,2p,3s,3p], H[1s,2s] & \ref{fig: Figure 1}c \\
\hline\hline
\end{tabular}
\end{table}
}

Metz et al ~\cite{metz2019molecular} and Li et al. \cite{Li2009} demonstrated that CCSD(T)/aug-cc-pVQZ calculations closely reproduce the results of the CCSD(T)/complete basis set (CBS) limit for the methane dimer. Metz et al.~\cite{metz2019molecular} also demonstrated this for water dimer. All of our simulations are therefore done with the aug-cc-pVQZ basis set. We simulate the water and methane dimers with the active spaces listed in Table I. 

\begin{figure}[ht]
     \centering
     \includegraphics[width=\textwidth]{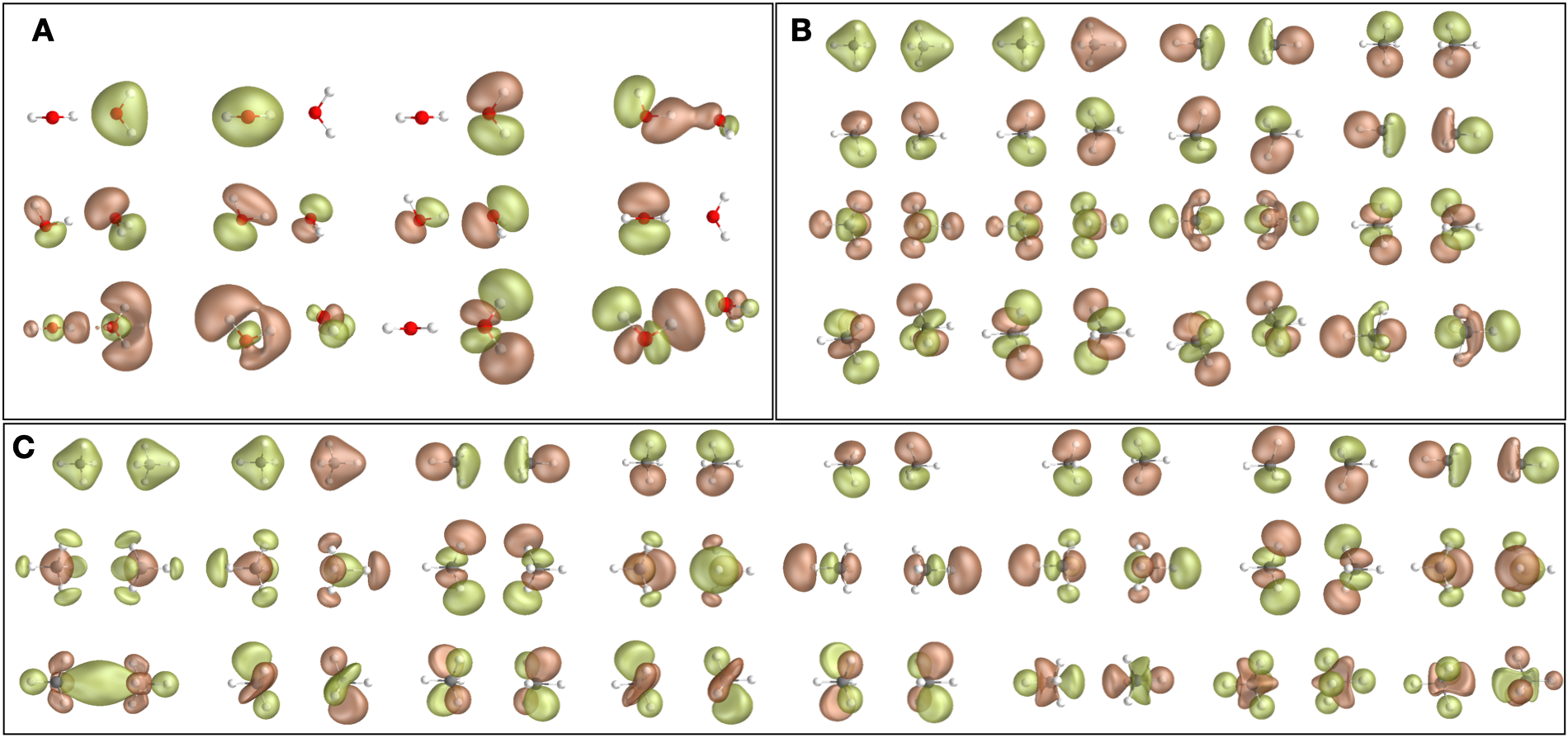}
     \caption{Active spaces used in this work: (A) (16e,12o) of the water dimer, (B) (16e,16o) of the methane dimer, (C) (16e,24o) of the methane dimer.}
     \label{fig: Figure 1}
\end{figure}

We construct these active spaces using the atomic valence active space (AVAS) method~\cite{Sayfutyarova2017} as implemented in the PySCF 2.6.2 software package~\cite{sun2020recent,sun2018pyscf,sun2015libcint}, and select active-space orbitals that overlap with the atomic orbitals (AOs) listed in column 3 of the table. The active-space orbitals of the water and methane dimers are shown in Fig.~\ref{fig: Figure 1}. Orbital visualization is performed with Pegamoid.\cite{galvan_pegamoid}

In each active space, we perform CCSD and CCSD(T) calculations with PySCF 2.6.2. For water and methane dimers we also perform CASCI(16e,12o) and CASCI(16e,16o) simulations, respectively, using PySCF 2.6.2. For the (16e,24o) active space of the methane dimer we perform HCI calculations with the SHCI-SCF 0.1 interface between PySCF 2.6.2 and DICE 1.0~\cite{holmes2016heat,smith2017cheap,sharma2017semistochastic}. Further details of HCI calculations can be found in the Supplementary Information. Along with active-space simulations, we perform complete CCSD and CCSD(T) calculations with ORCA 5.0.4~\cite{neese2022software}. The geometries of equilibrium structures of the water and methane dimer originate from works by Temelso et al~\cite{temelso2011benchmark} and by Rezac and Hobza~\cite{rezac2013describing}, sourced through the BEGDB database~\cite{rezac2008quantum}. We describe the generation of PES geometries for water and methane dimers in the Supplementary Information.

\subsection{Quantum Computing}

We start from the active-space Hamiltonian, written in second quantization as
\begin{equation}\label{eq3}
    \hat{H} = E_0 + \sum_{\substack{pr \\ \sigma}}h_{pr}\hat{a}_{p\sigma}^\dagger \hat{a}^{\phantom{\dagger}}_{r\sigma}+\sum_{\substack{prqs \\ \sigma\tau}} \frac{(pr|qs)}{2} \hat{a}_{p\sigma}^\dagger \hat{a}_{q\tau}^\dagger \hat{a}^{\phantom{\dagger}}_{s\tau} \hat{a}^{\phantom{\dagger}}_{r\sigma}
    \;,
\end{equation}
where $\hat{a}^\dagger$ ($\hat{a}$) are creation (annihilation) operators, $p$,$r$,$s$, and $q = 1 \dots M$ denote basis set element, $\sigma$ and $\tau$ denote spin-$z$ polarizations, $h_{pr}$ and $(pr|qs)$ are the one- and two-body electronic integrals, and $E_0$ is a constant accounting for the electrostatic interactions between nuclei and electrons in occupied inactive orbitals. We obtain the quantities $E_0$, $h_{pr}$, and $(pr|qs)$ for the selected active spaces using PySCF.

We prepare our wavefuntion guesses $|\Psi\rangle$, used to approximate the ground state of Eq.~\eqref{eq3}, from a truncated version of the local unitary cluster Jastrow (LUCJ) ansatz~\cite{motta2023bridging}
\begin{equation}
\label{eq4}
|\Psi\rangle=\prod_{\mu=0}^{L-1}e^{\hat{K}_\mu}e^{i\hat{J}_{\mu}}e^{-\hat{K}_\mu} 
| {\bf{x}}_{\mathrm{RHF}} \rangle \;,
\end{equation}
where ${\hat{K}_\mu}=\Sigma _{pr,\sigma} K^{\mu}_{pr}\hat{a}_{p\sigma}^\dagger\hat{a}^{\phantom{\dagger}}_{r\sigma}$ are one-body operators, ${\hat{J}_\mu}=\Sigma _{pr,\sigma \tau} J^{\mu}_{p\sigma, r\tau}\hat{n}_{p\sigma}\hat{n}_{r\tau}$ are suitable (vide infra) density-density operators, and $| {\bf{x}}_{\mathrm{RHF}} \rangle$ is the restricted closed-shell Hartree-Fock (RHF) state. We use the Jordan-Wigner (JW) transformation~\cite{jordan1993paulische} to map the fermionic wavefunction Eq.~\eqref{eq4} onto a qubit wavefunction that can be prepared executing a quantum circuit. 
The JW transformation maps the Fock space of fermions in $M$ spatial orbitals onto the Hilbert space of $2M$ qubits, where the basis state $|{\bf{x}} \rangle$ is parametrized by a bitstring ${\bf{x}} \in \{0,1\}^{2M}$ and represents an electronic configuration where the spin-orbital $p\sigma$ is occupied (empty) if $x_{p\sigma}=1$ ($x_{p\sigma}=0$). 
We prepare the wavefunction Eq.~\eqref{eq4} by executing the following quantum circuit: a single layer of Pauli-X gates prepares the basis state $| {\bf{x}}_{\mathrm{RHF}} \rangle$, a Bogoliubov circuit~\cite{aleksandrowicz2019qiskit} (with linear depth, quadratic number of gates, and a 1D qubit connectivity) encodes each orbital rotation $e^{ \pm \hat{K}_\mu}$, and a circuit of Pauli-ZZ rotations encodes each density-density interaction $e^{i\hat{J}_{\mu}}$. When $J^{\mu}$ is a dense matrix, Pauli-ZZ rotations are applied across all pair of qubits, requiring all-to-all qubit connectivity or a substantial overhead of swap gates. To mitigate these quantum hardware requirements LUCJ imposes a ``locality'' approximation, i.e., it assumes $J^{\mu}_{p\sigma, r\tau}=0$ for all pairs of spin-orbitals that are not mapped onto adjacent qubits under JW~\cite{motta2023bridging} (as a consequence, a circuit with constant depth and linear number of gates encodes each $e^{i\hat{J}_{\mu}}$ operator). Hence, the number of layers ($L-1$) in Eq.~\eqref{eq4} is formally equal to 1.5. As the result the specific form of $|\Psi\rangle$ used in this work is expressed as $|\Psi\rangle=e^{-\hat{K}_2} e^{\hat{K}_1}e^{i\hat{J}_{1}}e^{-\hat{K}_{1}} | {\bf{x}}_{\mathrm{RHF}} \rangle$.
We parametrize the LUCJ circuit based on amplitudes computed from classical restricted closed-shell CCSD within the given active space~\cite{robledo2024chemistry}, yet a further quantum-classical parameter optimization could further improve the quality of the ground-state approximation.
We produce the LUCJ circuits using the ffsim library~\cite{ffsim2024} interfaced with Qiskit 1.1.1~\cite{aleksandrowicz2019qiskit,javadi2024quantum}.

\begin{figure}[ht]
     \centering
     \includegraphics[width=1.0\textwidth]{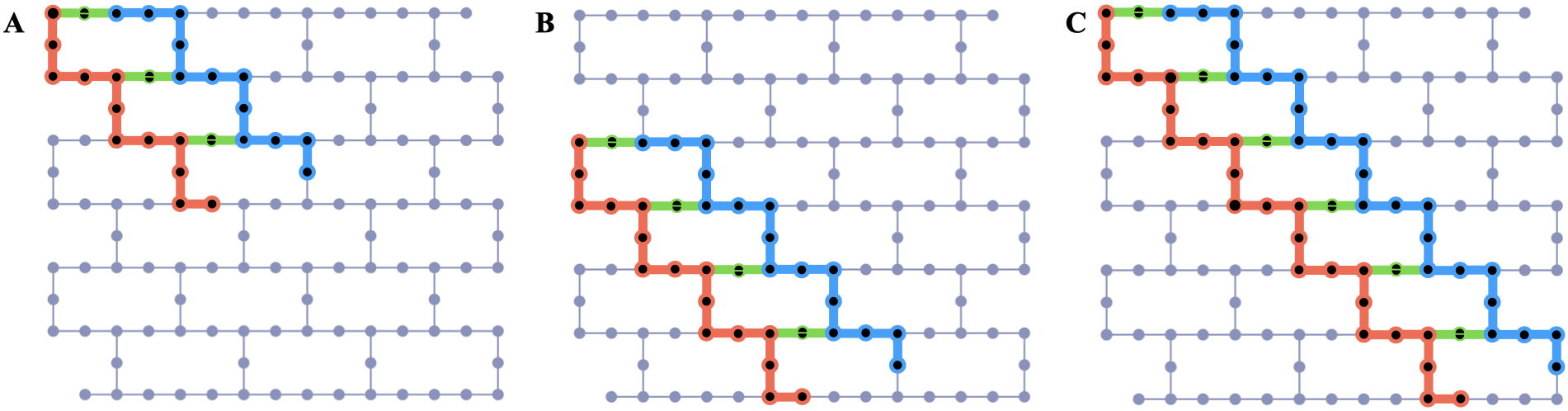}
     \caption{Qubit layouts of LUCJ circuits executed in this work: (A) (16e,12o) water dimer simulations using 27 qubits on ibm$\_$cleveland, (B) (16e,16o) methane dimer simulations using 36 qubits on ibm$\_$cleveland, and (C) (16e,24o) methane dimer simulations using 54 qubits on ibm$\_$kyiv. Qubits used to encode occupation numbers of $\alpha$ ($\beta$) spin-orbitals are shown in red (blue). Auxiliary qubits used to execute density-density interactions between $\alpha$ and $\beta$ spin-orbitals are marked in green.}
     \label{fig: Figure 2}
\end{figure}

The qubit layouts of the LUCJ circuits used for (16e,12o) water dimer, (16e,16o) methane dimer, and (16e,24o) methane dimer simulations are shown in Fig.~\ref{fig: Figure 2}a, \ref{fig: Figure 2}b, and \ref{fig: Figure 2}c, respectively. We execute these circuits on IBM's 127-qubit Eagle devices ibm$\_$cleveland and ibm$\_$kyiv. In all our quantum computing experiments, we used gate (not measurement) twirling over random 2-qubit Clifford gates~\cite{wallman2016noise} and dynamical decoupling~\cite{viola1998dynamical,kofman2001universal,biercuk2009optimized,niu2022effects} -- available through the SamplerV2 primitive of Qiskit's runtime library -- to mitigate quantum errors.

Upon executing the LUCJ circuits, we measure $|\Psi \rangle$ in the computational basis. Repeating this produces a set of measurement outcomes (or ``shots'') 
\begin{equation}\label{eq5}
\tilde{\chi} = \left \{ {\bf{x}} | {\bf{x}} \sim  \tilde{p}({\bf{x}}) \right \}
\end{equation}
in the form of bitstrings ${\bf{x}} \in \left \{0,1 \right \}^{2M}$, each representing an electronic configuration (Slater determinants) distributed according to $\tilde{p}({\bf{x}})$.
While on a noiseless device configurations are distributed according to $| \langle {\bf{x}} | \Psi \rangle|^2$, on a noisy device they follow a distribution $\tilde{p}({\bf{x}}) \neq | \langle {\bf{x}} | \Psi \rangle|^2$. In particular, $\tilde{p}({\bf{x}})$ breaks particle-number conservation and returns configurations with incorrect particle number. We use a technique called self-consistent configuration recovery~\cite{robledo2024chemistry}, executed on a classical computer, to restore particle-number conservation. The associated code is publicly available in the GitHub repository.~\cite{sqd_addon}
Within each step of self-consistent recovery, we sample $K$ subsets (or batches) of $\tilde{\chi}$ labeled $\tilde{\chi}_b$ with $b = 1 \dots K$. Each batch defines -- through a transformation~\cite{robledo2024chemistry} informed by an approximation to the ground-state occupation numbers $n_{p\sigma}$ -- a subspace $S^{(b)}$ of dimension $d$, in which we project the many-electron Hamiltonian as~\cite{kanno2023quantum,nakagawa2023adapt,robledo2024chemistry}
\begin{equation}
\label{eq6}
\hat{H}_{S^{(b)}}=\hat{P}_{S^{(b)}}\hat{H}\hat{P}_{S^{(b)}}
\;,
\end{equation}
where the projector $\hat{P}_{S^{(b)}}$ is
\begin{equation}
\label{eq7}
\hat{P}_{S^{(b)}} =\sum_{ {\bf{x}} \in S^{(b)}} | {\bf{x}} \rangle \langle {\bf{x}} |
\;.
\end{equation}
We compute the ground states and energies of the Hamiltonians in Eq.~\eqref{eq6}, $|\psi^{(b)} \rangle$ and $E^{(b)}$ respectively, and use the lowest energy across the batches, $\min_b E^{(b)}$, as the best approximation to the ground-state energy at the current iteration of the configuration recovery. We use the ground states $|\psi^{(b)} \rangle$ to obtain an updated set of occupation numbers,
\begin{equation}
\label{eq8}
n_{p\sigma}=\frac{1}{K} \sum_{1\leq b\leq K } \langle \psi^{(b)} | \hat{n}_{p\sigma} |\psi^{(b)} \rangle
\;,
\end{equation}
that we use in the next iteration of configuration recovery to produce the subspaces $S^{(b)}$. We repeat the iterations of self-consistent configuration recovery until convergence of the energy $\min_b E^{(b)}$. At the first iteration of self-consistent configuration recovery, we initialize $n_{p\sigma}$ from the  measurement outcomes in $\tilde{\chi}$ with the correct particle number. We summarize the details of our SQD calculations in Table II.

\begin{table}
\caption{Details of SQD calculations.}
\begin{tabular}{llllllll}
\hline\hline
species       & active space & $|\tilde{\chi}| \, [10^3]$ & $K$ & $|\tilde{\chi}_b|\, [10^3]$ & $d$ & CPUs, code & steps \\
\hline
water dimer    & (16e,12o) & $200$    & 10  & $10$    & $24.5\cdot10^4$ & 10, PySCF & 10 \\
methane dimer  & (16e,16o) & $200$    & 10  & $20$    & $12.6\cdot10^7$ & 10, PySCF & 10 \\
methane dimer  & (16e,24o) & $300$    & 4   & $8.5$   & $24.9\cdot10^7$ & 20, DICE  & 5  \\
\hline\hline
\end{tabular}
\end{table}

We demonstrate that for SQD (16e,16o) simulations of the methane dimer at 3.638 $\text{\AA}$ a $|\tilde{\chi}_b|= 20.0\cdot10^3$ is necessary to reach agreement within 0.010 kcal/mol when compared against CASCI (16e,16o). We show how the predicted total energies in these simulations improve with an increase of $|\tilde{\chi}_b|$ from $5.0\cdot10^3$ to $20.0\cdot10^3$ in Figure S3. We also demonstrate that in SQD(16e,16o) simulations the linear energy-variance relation allows for utilization of energy extrapolation which reproduces similar binding energies as simulation with $|\tilde{\chi}_b|$ = $20.0\cdot10^3$ while using substantially lower values of $|\tilde{\chi}_b|$. The extrapolation is done for the total energy of the dimer as the function of the Hamiltonian variance divided by the square of the variational energy, where Hamiltonian variance ($\Delta H$) is calculated as $\Delta H = \langle \psi^{(k)} | \hat{H}^2 | \psi^{(k)} \rangle - \langle \psi^{(k)} | \hat{H} | \psi^{(k)}  \rangle^2$. The extrapolation is done based on three points with $|\tilde{\chi}_b|$ of $9.0\cdot10^3$, $11.0\cdot10^3$, and $14.0\cdot10^3$, which allows for the reduction of the maximum $|\tilde{\chi}_b|$ by $6.0\cdot10^3$. This choice of values for $|\tilde{\chi}_b|$ allows for an even distribution of $\Delta H$ values used in extrapolation.  The extrapolated energies are compared against CASCI(16e,16o) simulations and SQD(16e,16o) simulations with $|\tilde{\chi}_b|$ = $20.0\cdot10^3$.
 
  We compute the ground-state eigenpairs of the Hamiltonians Eq.~\eqref{eq6} using the iterative Davidson method on 10 CPUs with PySCF's selected configuration interaction (SCI) solver for SQD (16e,12o) simulations of the water dimer and SQD (16e,16o) simulations of the methane dimer. We achieve parallelization across 10 CPUs with Ray 2.33.0~\cite{moritz2017ray} where the eigenstate solver within each of the 10 batches is using 1 CPU. For SQD (16e,24o) simulations of the methane dimer, we utilize the SCI solver of DICE and 20 CPUs, where the eigenstate solver within each of the 5 batches is using 4 CPUs. Further parallelization is possible with the SCI solver of DICE, as was demonstrated previously~\cite{robledo2024chemistry}. The SQD (16e,12o) simulations of the water dimer and SQD (16e,16o) simulations of the methane dimer are done for the distances that are described in the Supplemental Information while SQD (16e,24o) simulations of the methane dimer are only done for 3.638 $\text{\AA}$.
\section{Results}

\begin{figure}[hb]
     \centering
     \includegraphics[width=\textwidth]{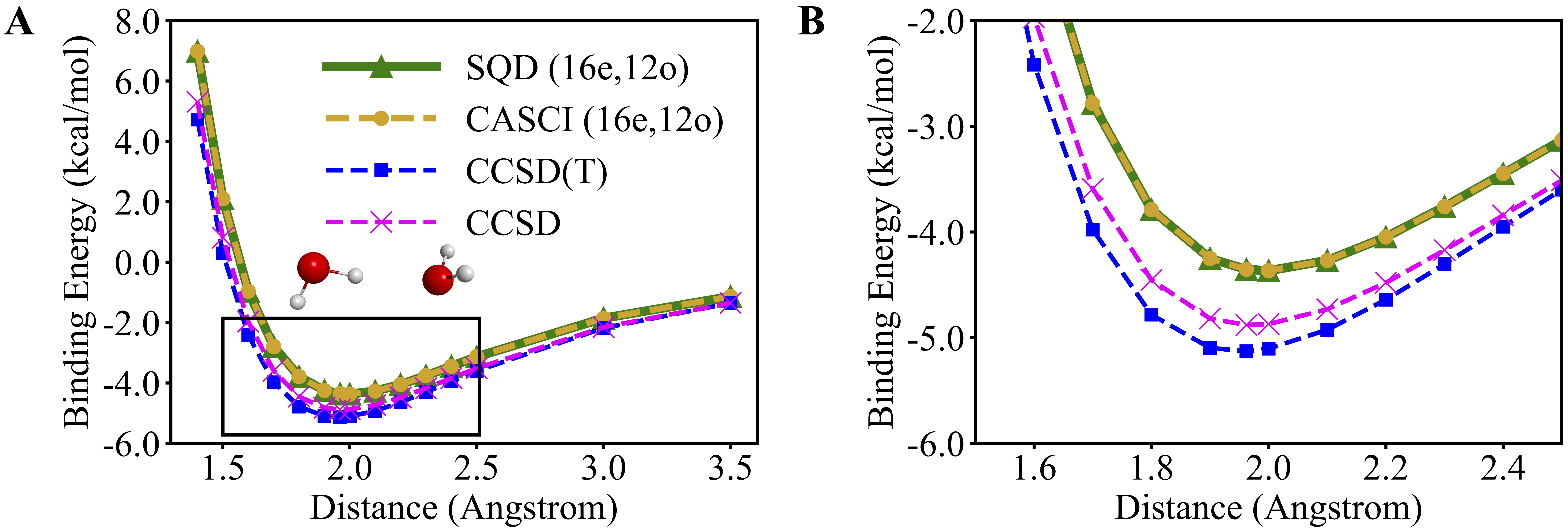}
     \caption{Binding energies of the water dimer along the PES, where distances between oxygen atoms range between 1.400 and 3.500 $\text{\AA}$. (A) the entire range of bondlengths, and (B) a magnified region near equilibrium, highlighted in panel (A) as a black box. Light brown, blue, and magenta dashed lines with circle, square, and cross markers depict the PES calculated with the CASCI (16e,12o), CCSD(T), and CCSD methods, respectively. The solid green line with triangular markers depicts the PES calculated with the SQD (16e,12o).}
     \label{fig: Figure 3}
\end{figure}

Figure~\ref{fig: Figure 3} shows the binding energy of the water dimer as a function of the oxygen-oxygen distance using SQD and CASCI. The SQD and CASCI potential energy surfaces closely align, deviating from each by less than 0.001 kcal/mol. This close alignment is an indication that both methods have accurately solved the Schodinger equation in the active space. The active-space SQD and CASCI calculations cannot capture dynamical correlation from inactive orbitals. To quantify the extent of the active-space approximation, we also compute the potential energy surface using CCSD and CCSD(T) in the full aug-cc-pVQZ basis.
The perturbative triples do not have a drastic effect on the binding energy between water monomers and the close agreement between CCSD and CCSD(T) calculations is shown in Figure 3b. The excellent agreement between CCSD and CCSD(T) in the full basis and between SQD and CASCI in the active space indicates that the differences between SQD and CCSD(T) are due to the active-space approximation underlying the former.
The CCSD(T) and SQD potential energy curves are in reasonable agreement with each other, the highest deviation being observed at 1.400 $\text{\AA}$ and corresponding to 2.263 kcal/mol. Despite this reasonable agreement and the ability of SQD to capture hydrogen bonding, there are quantitative differences in the predicted binding energies, -5.129 kcal/mol and -4.366 kcal/mol for CCSD(T) and SQD respectively, and the lowest-energy distances, 1.962 $\text{\AA}$ and 2.000 $\text{\AA}$ CCSD(T) and SQD respectively. The quantitative differences between SQD and CCSD and CCSD(T) shown in Fig. 3 are a consequence of SQD not being carried out in the full basis set. 

\begin{figure}[ht]
     \centering
     \includegraphics[width=\textwidth]{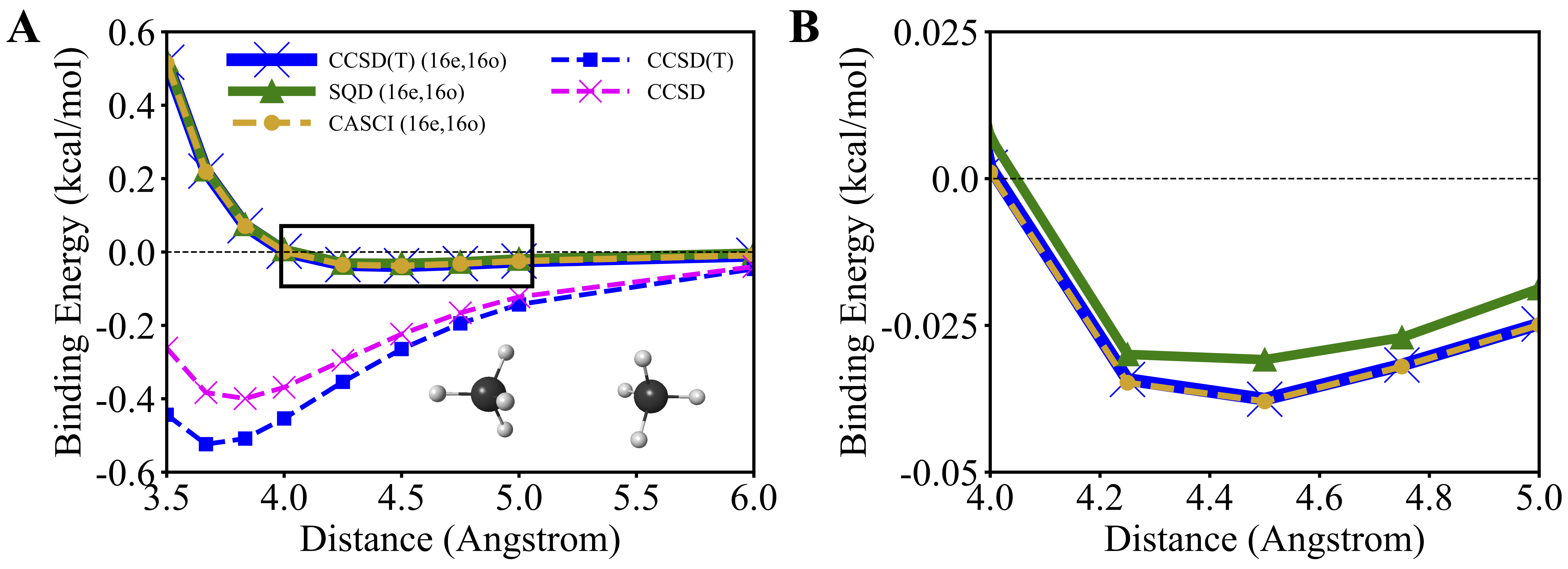}
     \caption{Binding energies of the methane dimer along the PES, where the distances between the carbon atoms range between 3.500 and 6.000 $\text{\AA}$.  Active-space simulations use (16e,16o). (A) the entire range of bondlengths, and (B) a magnified region near equilibrium, highlighted in panel (A) as a black box. Light brown, blue, and magenta dashed lines with circle, square, and cross markers depict the PES calculated with the CASCI (16e,16o), CCSD(T), and CCSD methods, respectively. The solid green line with triangular markers depicts the PES calculated with the SQD (16e,16o). The solid blue line represents CCSD(T) (16e,16o) calculations. The black horizontal dashed line indicates the zero value of the binding energy.}
     \label{fig: Figure 4}
\end{figure}

Figure~\ref{fig: Figure 4} shows the binding energy of the methane dimer -- with (16e,16o) active space for the dimer and monomer respectively -- as a function of the carbon-carbon distance using SQD and HCI. Figure~\ref{fig: Figure 4} focuses on the attractive region, whereas the full curve is shown in Figure S5 of the SI.
The SQD (16e,16o) and CASCI (16e,16o) data are closely aligned, with deviations below 0.005 kcal/mol. We interpret the excellent agreement between SQD (16e,16o) and CASCI (16e,16o) as an indication that the active-space Schrodinger equation is solved accurately. 
SQD (16e,16o) predicts the interaction between the monomers to be only marginally attractive, with a binding energy of -0.038 kcal/mol and a lowest-energy distance around 4.500 $\text{\AA}$. On the other hand, full-basis CCSD and CCSD(T) calculations predict binding energies of -0.399 kcal/mol and -0.524 kcal/mol, respectively, at distances 3.834 $\text{\AA}$ and 3.667 $\text{\AA}$, respectively. Despite some differences quantifying the importance of perturbative triple corrections, both full-basis calculations predict a substantially more pronounced tendency to binding than SQD (16e,16o) and CASCI (16e,16o). This is because, although SQD (16e,16o) and CASCI (16e,16o) calculations can accurately capture the active-space electronic correlation, they cannot account for the residual dynamical electron correlation, unlike full-basis CCSD and CCSD(T).

\begin{figure}[hb!]
     \centering
     \includegraphics[width=\textwidth]{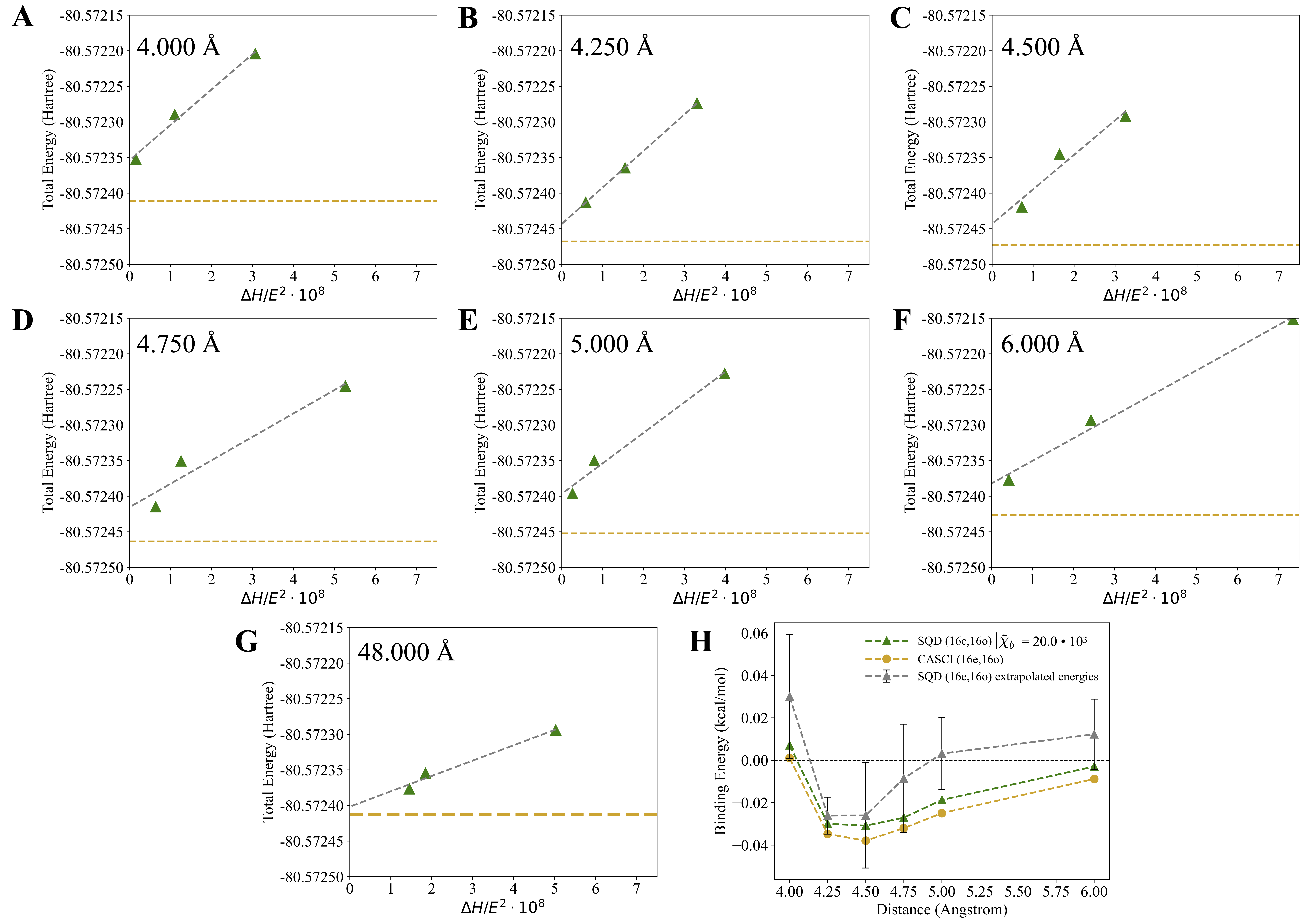}
     \caption{Extrapolated SQD (16e,16o) energies of the methane dimer along the PES, for the 4.000, 4.250, 4.500, 4.750, 5.000, and 6.000 $\text{\AA}$ distances between the carbon atoms. Extrapolations are done using three points with $|\tilde{\chi}_b|$ of $9.0\cdot10^3$, $11.0\cdot10^3$, and $14.0\cdot10^3$. Hamiltonian variance ($\Delta H$) is calculated as $\Delta H = \langle \psi^{(k)} | \hat{H}^2 | \psi^{(k)} \rangle - \langle \psi^{(k)} | \hat{H} | \psi^{(k)}  \rangle^2$. (A) - (F) total energy extrapolations for methane dimer 4.000, 4.250, 4.500, 4.750, 5.000, and 6.000 $\text{\AA}$ distances, (G) total energy extrapolations at 48.000 $\text{\AA}$ distance, and (H) binding energy in methane dimer calculated with extrapolated SQD (16e,16o) total energies compared against CASCI (16e,16o) simulations and SQD (16e,16o) simulations with $|\tilde{\chi}_b|$ = $20.0\cdot10^3$. Green triangles and green dashed lines indicate SQD (16e,16o) energies calculated with $|\tilde{\chi}_b|$ = $20.0\cdot10^3$. Grey triangles and grey dashed lines indicate extrapolated SQD (16e,16o) energies. Light brown circles and dashed lines indicate CASCI (16e,16o) energies. Black horizontal dashed line indicates the zero value of the binding energy. Error bars indicate magnitude of error estimate in extrapolation.}
     \label{fig: Figure 5}
\end{figure}

Before proceeding with the expansion of the active space we first demonstrate that accurate SQD (16e,16o) calculations can be achieved with a reduced number of samples through the extrapolation of the total energies. The exact SQD (16e,16o) calculations require $|\tilde{\chi}_b|$ = $20.0\cdot10^3$ while the extrapolation is done based on three points with $|\tilde{\chi}_b|$ of $9.0\cdot10^3$, $11.0\cdot10^3$, and $14.0\cdot10^3$. Hence, the extrapolation allows for the reduction of the maximum required $|\tilde{\chi}_b|$ by $6.0\cdot10^3$. We show the SQD (16e,16o) total energy extrapolations for 4.000, 4.250, 4.500, 4.750, 5.000, and 6.000 $\text{\AA}$ distances in Figure 5a-f, while the extrapolation for the 48.000 $\text{\AA}$ distance is shown in Figure 5g. The resulting binding energies of the methane dimer are shown in Figure 5h and compared against the CASCI (16e,16o) simulations and SQD (16e,16o) simulations with $|\tilde{\chi}_b|$ = $20.0\cdot10^3$. Figure 5g shows that the extrapolated SQD (16e,16o) energies predict a binding energy in good qualitative agreement with exact SQD (16e,16o) simulations and CASCI (16e,16o). This result is promising for future simulations with large active spaces, where classical post-processing of SQD data becomes computationally expensive.

\begin{figure}[!ht]
     \centering
     \includegraphics[width=0.95\textwidth]{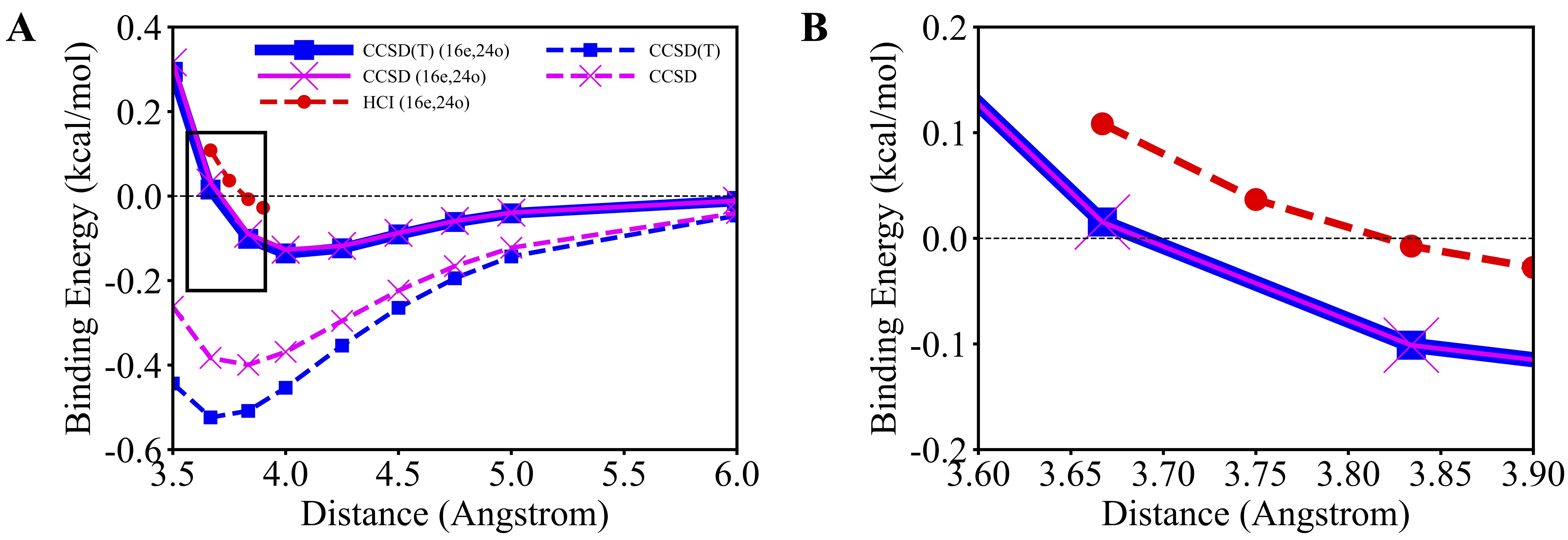}
     \caption{Binding energies of the methane dimer along the PES, where the distances between the carbon atoms range between 3.500 and 6.000 $\text{\AA}$. Active-space simulations use (16e,24o) and are performed over (A) the entire range of bondlengths, and (B) a magnified region near equilibrium, highlighted in panel (A) as a black box. Blue and magenta dashed lines depict PES calculated with CCSD(T) and CCSD methods, respectively. The dashed red line depicts the HCI (16e,24o) results. The solid blue and magenta lines represents CCSD(T) (16e,24o) and CCSD (16e,24o) calculations, respectively. Black horizontal dashed line indicates the zero value of the binding energy.}
     \label{fig: Figure 6}
\end{figure}

Next we analyze the effect of extending the active space on the predicted binding energy via the inclusion of virtual orbitals with carbon 3s and 3p character. First, in Fig.~\ref{fig: Figure 6}, we explore the performance of HCI in this extended (16e,24o) active space. Here, HCI is used in place of CASCI due to the fact that the (16e,24o) active space is prohibitively expensive in conventional CASCI simulations.
Figure~\ref{fig: Figure 6}a shows active-space calculations with HCI, CCSD, and CCSD(T).
The CCSD(T) (16e,24o) curve, in good agreement with CCSD (16e,24o), is substantially more attractive than in the (16e,16o) active space, predicting a binding energy of -0.136 kcal/mol at 4.000 $\text{\AA}$.
The size of the (16e,24o) active space prevents us from significantly lowering the parameter $\varepsilon_1$ which results in the underestimation of the total energy in HCI (16e,24o) calculations. In particular, at 3.834 {\AA} distance the HCI (16e,24o) calculations underestimate the binding energy by 0.094 kcal/mol comparing to CCSD(T) (16e,24o), as visible in Fig.~\ref{fig: Figure 6}b. Note that HCI (16e,24o) calculations were carried out over four distances (3.667, 3.750, 3.834, and 3.900 $\text{\AA}$). Within the target accuracy of the present work the prediction of HCI (16e,24o) binding energies for other geometries around the CCSD(T) (16e,24o) minimum is dramatically more computationally expensive.

\begin{figure}[ht]
     \centering
     \includegraphics[width=0.5\textwidth]{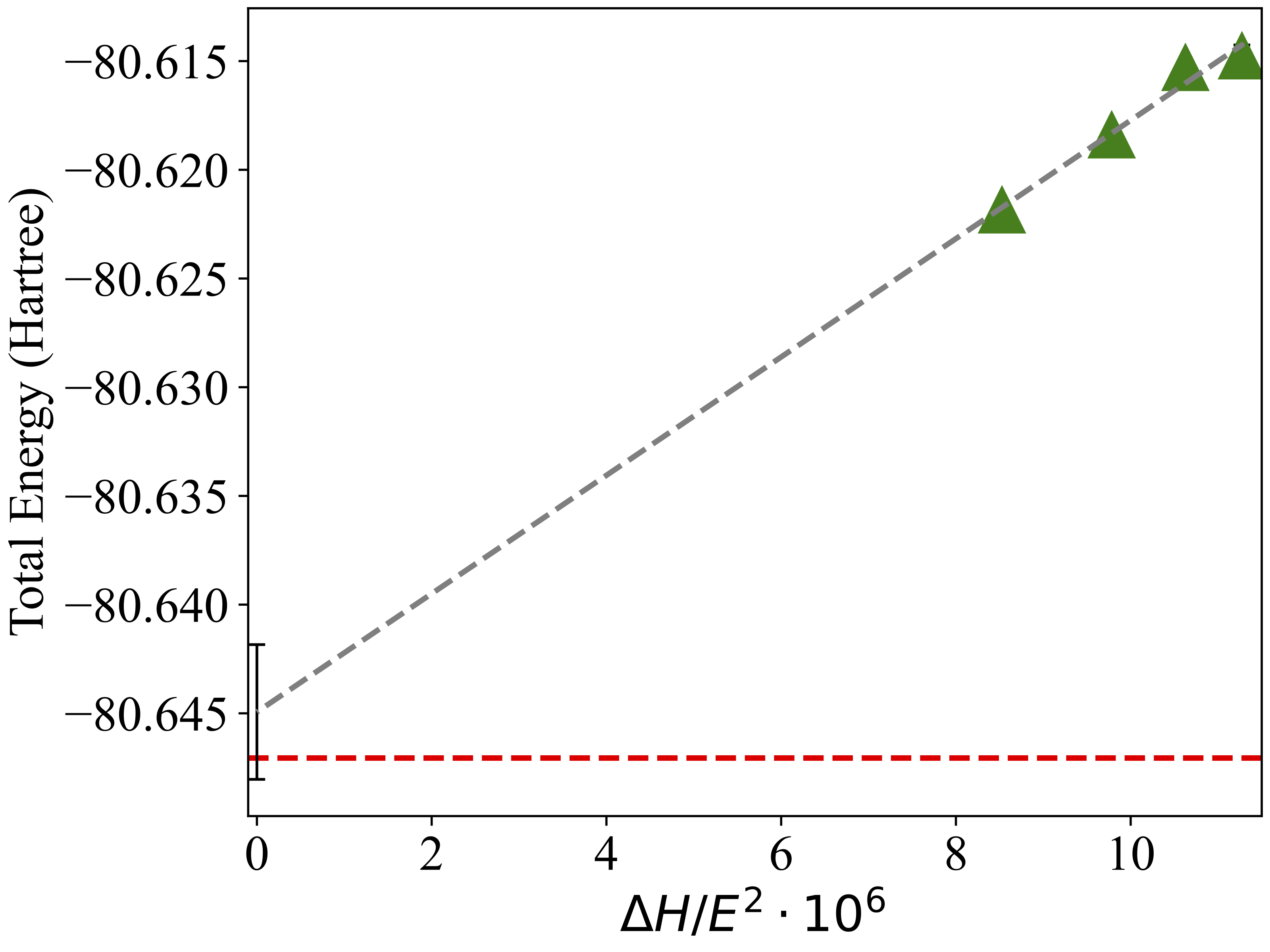}
     \caption{Extrapolation of SQD (16e,24o) total energies for the methane dimer at 3.638 $\text{\AA}$ distance between the carbon atoms. Extrapolations are done using four points with $|\tilde{\chi}_b|$ of $5.5\cdot10^3$, $6.5\cdot10^3$, $7.5\cdot10^3$, and $8.5\cdot10^3$. Hamiltonian variance ($\Delta H$) is calculated as $\Delta H = \langle \psi^{(k)} | \hat{H}^2 | \psi^{(k)} \rangle - \langle \psi^{(k)} | \hat{H} | \psi^{(k)}  \rangle^2$. Green triangles indicate SQD (16e,24o) energies. Grey dashed lines indicate extrapolated SQD (16e,24o) energies. Dashed red line indicates HCI (16e,24o) energies. Error bars indicate magnitude of error estimate in extrapolation.}
     \label{fig: Figure 7}
\end{figure}

We show the decrease in the SQD (16e,24o) total energy for the methane dimer at 3.638 $\text{\AA}$ with the increase of $|\widetilde{\chi}_b|$ from $5.0\cdot10^3$ to $8.5\cdot10^3$ in Fig.~\ref{fig: Figure 7}. The differences between the total energies predicted with SQD (16e,24o) and HCI (16e,24o) reduce from 31.8 milliHartree to 25.2 milliHartree when the $|\widetilde{\chi}_s|$ is increased from $5.0\cdot10^3$ to $8.5\cdot10^3$. The extrapolated total energy based on SQD (16e,24o) simulations with $|\tilde{\chi}_b|$ of $5.5\cdot10^3$, $6.5\cdot10^3$, $7.5\cdot10^3$, and $8.5\cdot10^3$ agrees with HCI (16e,24o) results within 2.12 milliHartree. Magnitude of the error estimate in extrapolation is $\pm$ 3.10 milliHartree. We believe that a further increase in the number of samples will allow us to advance the accuracy of SQD (16e,24o) calculations. To make SQD (16e,24o) calculations of the methane dimer more computationally feasible we are currently exploring parallelization options for calculations on this system as well as the analysis of the configurations with low contributions to the total energies.
\newpage
\section{Conclusions and Outlook}

We have presented quantum-centric simulations of the water and methane dimers using a sample-based quantum diagonalization method on IBM's Eagle quantum processors. This demonstration is a first simulation of noncovalent supramolecular interactions on quantum processors. The accuracy of SQD and HCI predictions of noncovalent interactions can be systematically improved by the addition of extended shells of virtual orbitals. We anticipate that further expansion of the active space through the inclusion of the virtual orbitals corresponding to the 3d shell of the heavy atoms will allow for an even more accurate description of non-covalent interactions with SQD and HCI, which will be the subject of future studies on quantum processors. Importantly, the present study lays out a framework for electronic structure calculations of noncovalent interactions on quantum hardware. 

Our findings demonstrate that SQD is capable of capturing noncovalent interactions between molecules at the level of theory chosen, with potential energy surfaces that closely align with those obtained through classical computational methods. We examine the binding energies of the water and methane dimers by comparing SQD with an analogous classical method, namely HCI. We also compare SQD against the CCSD(T) method, which is considered the gold standard for calculations of binding energies~\cite{carsky2010recent}. This comparison aims to evaluate the accuracy of SQD and to understand how the nature of the PES changes with different active-space selections. The ability of HCI and SQD to recover the dispersion interaction is highly dependent on the size and nature of the active space, which is especially critical for predicting the binding energy of the methane dimer. In fact, a previous study by Hapka et al.\cite{hapka2020dispersion} demonstrated that the ability of supramolecular multiconfigurational interaction calculations to recover the dispersion energy depends on the size of the active space and can be improved with systematic expansion of the active space.

The results obtained here demonstrate the improvements both in terms of accuracy and scale of quantum computations on chemical problems, enabling, on current quantum processors, use cases previously thought to belong to the fault-tolerant domain, such as the largest active space considered here for methane, which has ~1.3M Pauli operators. Further examples of problems that coule be enabled by our approach include quantum computing simulations of chemical reactivity of $CO_2$-fixating ruthenium catalyst proposed by Burg et al.~\cite{von2021quantum}, Ibrutinib drug simulations proposed by Blunt et al.\cite{blunt2022}, and the drug-discovery workflows proposed by Pyrkov et al.\cite{PYRKOV2023103675} and Kumar et al.\cite{10466774}, as well as multiple stages of drug optimization as described by Bonde et al.\cite{Bondechapter7} The SQD method allows for simulations of systems with qubit counts that is essential for projection-based embedding algorithm proposed by Ralli et al.\cite{Ralli2024} and can enhance the viability of fragment-based quantum computing simulations. Previously, VQE-based fragment molecular orbital (FMO),\cite{Lim2024} divide and conquer (DC),\cite{Yoshikawa2022} and density matrix embedding theory (DMET)\cite{kawashima2021optimizing,iijima2023accurate} simulations were limited to very simple illustrative systems. Shang et al. proposed a DMET-based massively parallel quantum computing approach based on VQE, but execution of their methodology was only possible on a quantum simulator rather than actual hardware.\cite{Shang2023}

Such limitations in fragment-based VQE simulations are due to the fact that the number of orbitals that could be described with reasonable accuracy on actual hardware in the VQE formalism within each fragment is very limited. Fragment-based simulations with SQD would allow for substantially higher number of orbitals in each individual fragment, making the quantum computing simulations of proteins and drug molecules possible.

In conclusion, combining quantum and classical computational resources in workflows like SQD opens the way for the use of current and near-future quantum technology to tackle computational challenges in small-molecule conformational search, drug-protein interactions and drug discovery.
\acknowledgements

The authors gratefully acknowledge financial support from the National Science Foundation (NSF) through CSSI Frameworks Grant OAC-2209717 and from the National Institutes of Health (Grant Numbers GM130641).


\bibliography{Supramolecular_SQD}

\clearpage
\setcounter{equation}{0}
\setcounter{section}{0}  
\setcounter{figure}{0}   
\renewcommand{\theequation}{S\arabic{equation}}
\renewcommand{\thefigure}{S\arabic{figure}}

\setcounter{page}{1}

{\section*{Supplementary Information: Accurate quantum-centric simulations of supramolecular interactions}}


\section{Geometries of potential energy surfaces}

The PES for the water dimer is calculated for distances between two oxygen atoms ranging between 1.400 and 3.500 $\text{\AA}$. We distribute the points for water dimer PES as 1.400, 1.500, 1.600, 1.700, 1.800, 1.900, 1.962, 2.000, 2.100, 2.200, 2.300, 2.400, 2.500, 3.000, and 3.500 $\text{\AA}$. All of water dimer simulations are done for all of these distances. The PES for the methane dimer is calculated for the distances between two carbon atoms ranging between 2.500 and 6.000 $\text{\AA}$. We distribute the points for methane dimer PES as 2.500, 2.750, 3.000, 3.167, 3.334, 3.500, 3.667, 3.834, 4.000, 4.250, 4.500, 4.750, 5.000, and 6.000 $\text{\AA}$. To calculate the total energy of unbound dimer we utilize the distance of 48.000 $\text{\AA}$ for both water and methane dimers. All of CASCI (16e,16o), CCSD, CCSD(T), CCSD (16e,16o), and CCSD(T) (16e,16o) simulations of methane dimer as well as SQD (16e,16o) simulations with $|\tilde{\chi}_b|$ = $20.0\cdot10^3$ are done for all of the distances described earlier. The methane dimer CASCI (16e,16o) simulations and SQD (16e,16o) simulations with $|\tilde{\chi}_b|$ = $20.0\cdot10^3$ are also performed for an additional distance of 3.638 $\text{\AA}$. The SQD (16e,16o) energy extrapolations using $|\tilde{\chi}_b|$ of $9.0\cdot10^3$, $11.0\cdot10^3$, and $14.0\cdot10^3$ are done for 4.000, 4.250, 4.500, 4.750, 5.000, 6.000, and 48.000 $\text{\AA}$ distances. In the case of HCI (16e,24o) simulations of the methane dimer, we use only distances of 3.638, 3.667, 3.750, 3.834, and 3.900 $\text{\AA}$. In case of SQD (16e,24o) simulations we only use the distance of 3.638 $\text{\AA}$. All calculations are done as single-point energy calculations with no geometry optimizations. To produce the geometries studies in this work, we start from the equilibrium geometries and change the distance between the centers of the monomers, with the geometries of the individual monomers fixed.

\section{Details of HCI calculations}

In HCI simulations instead of generating all of the single and double excitations one generates only those single and double excitations that correspond to Hamiltonian matrix elements exceeding a threshold $\varepsilon$. In HCI $\varepsilon_1$ controls which determinants will be included in the variational wave function.   In our HCI calculations, we use values of $\varepsilon_1$ equal to $5\cdot10^{-6}$ (during initial variational steps) and $1\cdot10^{-6}$ (during later variational steps). We do not use the non-variational perturbative correction in our HCI calculations. Hence, our HCI calculations are fully variational, which allows for more appropriate comparison between the SQD and HCI results.

\section{Effect of number of samples on total energy in SQD(16e,16o)}
\begin{figure}[ht!]
     \centering
     \includegraphics[width=\textwidth]{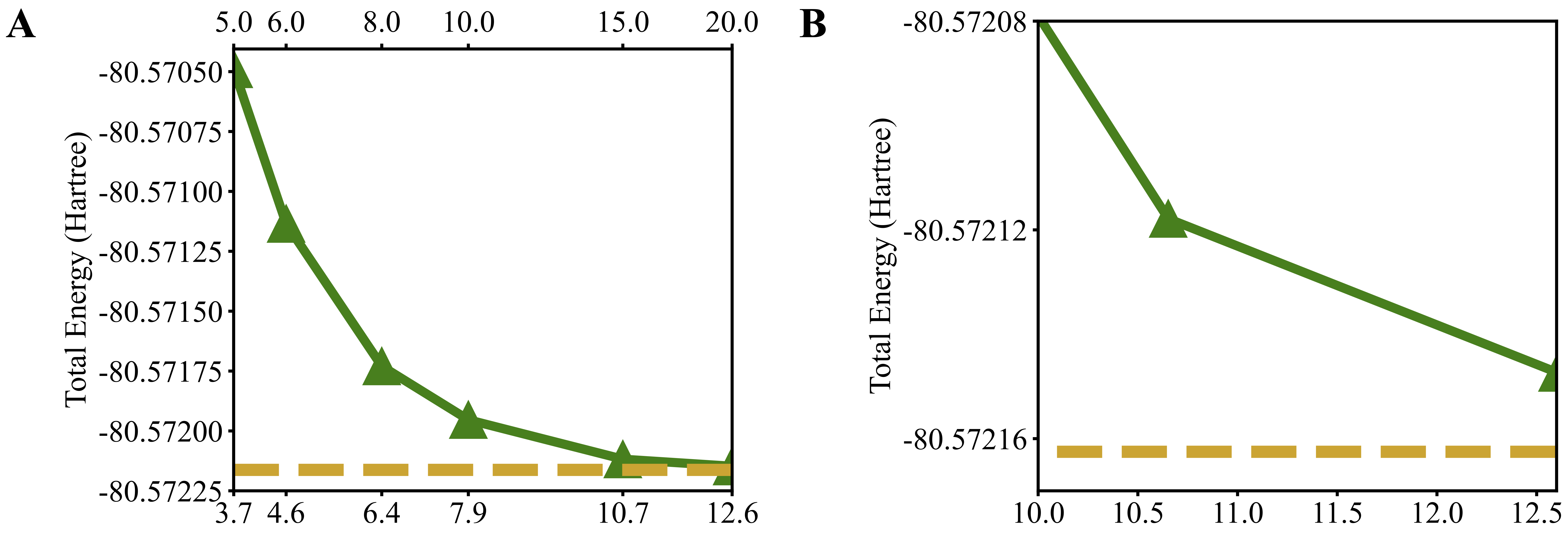}
     \caption{Total energy of methane dimer predicted with SQD (16e,16o) at 3.638 $\text{\AA}$ distance between the monomers as the function of $d \cdot 10^7$. (A) the entire range of \textit{d}, and (B) a magnified region with largest values of \textit{d}, highlighted in panel (A) as a black box. The secondary x-axis demonstrates the value of $|\widetilde{\chi}_b| \cdot 10^3$ producing the given value of $d \cdot 10^7$. Solid green line with triangular markers shows SQD (16e,16o) results. Horizontal dashed light brown line indicates the total energy from CASCI (16e,16o) calculation.}
     \label{fig: Figure S3}
\end{figure}

\newpage
\section{PES of methane dimer including the repulsive region}
\begin{figure}[ht!]
     \centering
     \includegraphics[width=0.75\textwidth]{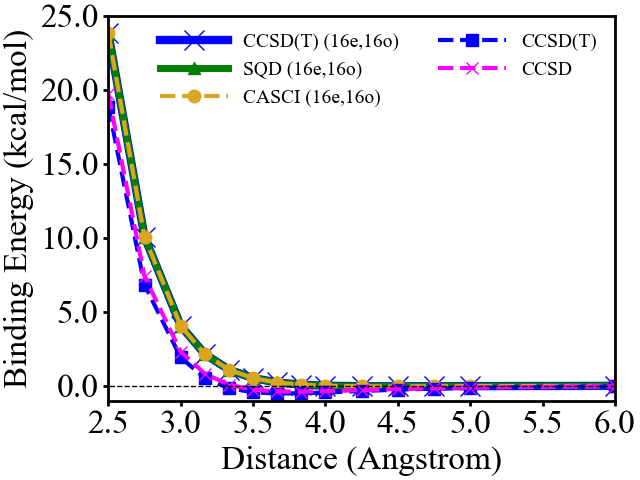}
     \caption{Binding energies of the methane dimer along its PES, where the distances between the centers of methane molecules range between 2.500 and 6.000 Å. Active space simulations are performed with (16e,16o). Light brown, orange, and blue dashed lines with circle markers depict PES calculated with CASCI, CCSD, and CCSD(T) methods, respectively. Solid yellow line with triangular markers depicts the PES calculated with the SQD method. Solid blue line represents CCSD(T) calculations using an active space. Black horizontal dashed line indicates the zero value of the binding energy.}
     \label{fig: Figure S6}
\end{figure}

\end{document}